\documentclass[12pt]{article}
\usepackage{graphicx}

\catcode`\@=11
\@addtoreset{equation}{section}
\catcode`\@=12

\newcommand{\bea}{\begin{eqnarray}}
\newcommand{\eea}{\end{eqnarray}}

\newcommand{\TPM}{'t Hooft-Polyakov monopole}

\def\be{\begin{equation}}
\def\ee{\end{equation}}

\def\fr{\frac}
\def\a{\alpha}
\def\b{\beta}

\def\d{\delta}
\def\e{\epsilon}

\def\l{\lambda}

\def\th{\theta}
\def\w{\omega}

\def\d{\delta}

\def\p{\partial}

\def\nn{\noindent}

\title{{\small\hfill IMSc/2002/12/42}\\ 
\textbf{Half-monopoles and half-vortices in the Yang-Mills theory}}
\author{E. Harikumar\footnote{hari@imsc.res.in},
Indrajit Mitra\footnote{indrajit@imsc.res.in}~ and
H. S. Sharatchandra\footnote{sharat@imsc.res.in} \\\\
The Institute of Mathematical Sciences,\\ C.I.T. Campus, Taramani P.O.,\\
Chennai 600 113, India}
\date{}
\begin{document}
\maketitle
\begin{abstract}
It is demonstrated that there are smooth Yang-Mills potentials
which correspond to monopoles and vortices of one-half winding number.
They are the generic configurations,
in contrast to the integral winding number configurations like the
't Hooft-Polyakov monopole.
\end{abstract}
\nn Keywords: Monopole, Poincar\'e-Hopf index, one-half winding number\\
\nn PACS numbers: 14.80.Hv, 11.15.-q, 11.15.Tk\\
\newpage
In this letter, we demonstrate Yang-Mills field configurations of monopoles and vortices
with half the usual charges. We show
that these are the generic field configurations,
in contrast to the integral winding number configurations such as the
't Hooft-Polyakov monopole.
In Refs.\ \cite{rps} and \cite{his}, the
monopole configuration was related to the singularities of the
eigenvector fields of the real symmetric matrix
\be
S_{ij}(x)=B_{i}^a(x)B_{j}^a(x),
\label{sij}
\ee
where $B_{i}^a=\e_{ijk}(\p_j A_{k}^a-\fr{1}{2}\e^{abc}A_{j}^bA_{k}^c)$
is the $SO(3)$ magnetic field. Such singularities arise due to indeterminacy
of the directions of the eigenvectors, and so it is crucial that the
eigenvalues become degenerate at the points
of singularity \cite{th}. The topology of the configuration can be
traced to these singularities. We refer to these points of
singularities as the `centres' of the topological configurations. For
the \TPM~\cite{tpm, goddard},
\be
S_{ij}=\a(r^2)\d_{ij} +\b(r^2) x_i x_j
\label{stpm}
\ee
where $\a$ and $\b$ are functions of the distance $r$ from the origin
only.  One of the eigenfunctions is the radial vector $ x_i$, with
unit winding number.
This has indeterminate direction at the origin $r=0$. But there is no
contradiction because $S~\propto~I$ (the identity matrix) at the
origin, and any vector is an eigenvector.

Note that the entries of the matrix $S$ are smooth
in the co-ordinates $x_i$
at the origin. Singularities arise in spite of this, due to the
eigenvalue equation.

The \TPM~has some exceptional features which are not generic. The first of
these is that two eigenvalues are degenerate everywhere. Secondly, the
entries of the matrix $S$ are quadratic in the co-ordinates. Thus, in
the Taylor series expansion of $S_{ij}(x)$ about the origin, linear
terms are missing. Both these features are a consequence of the
rotational invariance of the \TPM. (This rotational invariance is under
simultaneous and equal rotations in physical and isospin spaces.)

In this letter, we analyse the generic case, i.e., {\it we consider $S_{ij}(x)$
with linear terms in the Taylor expansion about the origin.} We find the novel
feature of half-integral winding number configurations and obtain the
interpretation of such configurations.

As we are interested in the eigenfunctions, we may appropriately
subtract a multiple of the identity matrix from $S_{ij}$. Also an
overall scale is irrelevant. {\it We will refer to the matrix after these
changes as $T_{ij}$.}

We first illustrate the possibility and meaning of configurations with half a
winding number using a $2\times 2$
real symmetric matrix field $T_{ij}(x,y)$. The paradigm is provided
by the matrix
\be
T=
\left(\begin{array}{cc}
x&y\\
y&-x
\end{array}\right)\,.
\label{tbyt}
\ee 
The eigenvalues are $\l_{\pm}= \pm r$, where $r=\sqrt{ x^2 + y^2}$. We
denote the corresponding eigenfunctions by $\zeta_i^\pm$.
The eigenfunction 
$\left(\begin{array}{c}\zeta_1^+\\\zeta_2^+\end{array}\right)$
has ${\zeta_1^+}/{\zeta_2^+}=y/(r-x)$. Thus the normalised
eigenfunction has the simple form
\be
\left(\begin{array}{c} \zeta_{1}^+\\\zeta_{2}^+\end{array}\right)=
\left(\begin{array}{c} \cos{\fr{\displaystyle\th}{\displaystyle 2}}
\\\sin{\fr{\displaystyle\th}{\displaystyle 2}}\end{array}\right)
\label{efhalfth}
\ee
 in the polar co-ordinates. Here $\theta=\tan^{-1}(y/x)$.
 
\begin{figure}
\begin{center}
\includegraphics[scale=0.25]{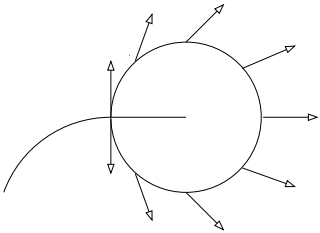}
\end{center}
\caption{A winding number half configuration: $\zeta_i^\pm$ changes sign
  when taken around any closed path enclosing the centre. 
  The curved line represents the (arbitrary) line
  of discontinuity.}
\label{fighalf}
\end{figure}

 The occurrence of half the polar angle in (\ref{efhalfth}) is
 significant. If we go round the origin once, the eigenvector changes
 the sign. It is not possible to define the vector field $\zeta_{i}^+(x)$
 continuously everywhere. There is necessarily a discontinuity (change
 of sign) across a ``branch cut" starting from the origin.  The
 choice of this branch cut is arbitrary, except that it starts at
 the origin. If we consider the complex vector
 $\zeta_1^+ + i \zeta_2^+= \exp(i\th/2)$, 
 the phase changes by $\pi$ when we go around the origin once. 
 In this sense, the winding number is half.
 We call this configuration a half-vortex.
 It can be
 checked that such a phase change takes place for the other eigenvector 
 $\zeta_i^-(x)$ too (Fig.\ \ref{fighalf}).

 We emphasise that the entries of the matrix $T_{ij}$ are smooth
 even at the origin.  In spite of this, the eigenvalue equation gave a
 discontinuous eigenvector field.
 
 It is easy to see that only half-integral winding number is possible
 in this case. The eigenvector of a real symmetric matrix is real and
 hence a non-degenerate eigenvector, after normalisation, is ambiguous only up to a sign.
 Therefore, when taken continuously around a closed path, the
 only possible change in the
 eigenvector on return to the initial
 point is by an overall sign. 
 This indeed happens in the present case.

 We now argue that this describes the situation in the generic case
 too.  The most general $2\times 2$ real symmetric linear in the 
 co-ordinates is
\be
 T =\left(\begin{array}{cc}
ax + by&cx + dy\\
cx + dy&ex +fy\\
\end{array}\right)\,.
\ee 
For the eigenvalue problem, we can subtract a multiple of identity
 matrix from $T$ given above. Subtracting $\fr{1}{2}((a+e)x +
(b+f)y)I$, we get a symmetric matrix.  We now choose the oblique
system of co-ordinates
\be 
2x^\prime=(ax +by)-(ex+fy),~~~~~y^\prime= cx + dy.
\ee 
(In the generic case these are linearly independent and a valid
choice of new co-ordinates.) With this we are back to the paradigm
considered in (\ref{tbyt}).

Therefore, on considering the Taylor series expansion of the entries $T_{ij}$
about the point of degeneracy, say $x=0$, $y=0$, it is clear that so long as the terms
linear in $x$ and $y$ are present, we get the phenomenon of one-half
winding number described above.

The situation will be totally different in the case where the entries
are quadratic in the co-ordinates. The simplest example is the one analogous
to the case of \TPM:
\be
T_{ij}= x_i x_j.
\label{tpm}
\ee 
Now the eigenvectors are 
$\hat r$=$\left(\begin{array}{c}
    \cos\th\\\sin\th\end{array}\right)$ and 
$\hat\th$=$\left(\begin{array}{c} -\sin\th\\\cos\th\end{array}\right)$.
For both of these eigenvectors, the winding number around the origin
is one, and the vector fields can be defined continuously everywhere
(except for the singularity at the origin).

\begin{figure}
\begin{center}
\includegraphics[scale=0.2]{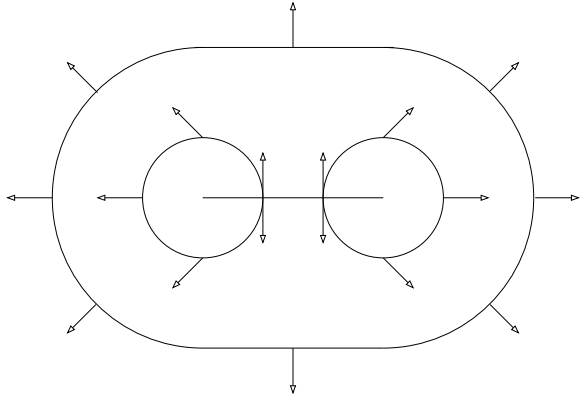}
\end{center}
\caption{Two winding number half configurations give a winding number one
  configuration at large distances.}
\label{fig2half}
\end{figure}

It is interesting to consider the case where
\be 
T= \left(
\begin{array}{cc}
x^2-a^2&xy\\
xy&y^2
\end{array}\right).
\label{tpd}
\ee
Now the double degeneracy is at two points, viz. $x_0=\pm a,
y_0=0$. Around each point the Taylor series expansion has the form
\be
T=  \left(
\begin{array}{cc}
2X&Y\\
~Y&0
\end{array}\right) + {\rm ~higher~order~terms~in~X,~Y}
                                                       \label{tbtho}
\ee
where $X=x_0(x- x_0)$ and $Y=x_0(y-y_0)$. The leading term has precisely
the form of the paradigm we considered (up to a multiple of the 
identity matrix). So
we get half a winding number around each point of degeneracy. We may
conveniently choose the line joining the two centres as the branch
cut.  The winding number along a curve enclosing both centres is one.
Indeed, as $a\to 0$, we recover from (\ref{tpd}) the winding number one configuration
considered in (\ref{tpm}). In this limit of $a\to 0$ , the pair of half winding
number configurations merge together to give winding number one
configuration (Fig.\ \ref{fig2half}).

Let us also consider the matrix
\be T=
\left(\begin{array}{cc}
x^2-a^2&ay\\
ay&0\end{array}\right).
\ee
This again has the same two points of degeneracy as the matrix in
(\ref{tpd}). However, in the present case the winding numbers are $\pm
1/2$ respectively. (The configuration around $(-a,0)$ is related to
our paradigm in (\ref{tbtho}) by reflection about the $X$-axis:
$Y\to -Y$. So it has the winding number $-1/2$.)  In the limit $a \to
0$, the eigenvectors are now $\left
  (\begin{array}{c}1\\0\end{array}\right)$ and $\left
  (\begin{array}{c}0\\1\end{array}\right)$, and each of them has a vanishing
winding number.

We have considered $2\times 2$ matrices though the  $S_{ij}$ relevant
for the Yang-Mills theory are $3\times 3$ matrices.
We regard the matrix in (\ref{tbyt}) as a block of the $3\times 3$ matrix
\be
T=\left(\begin{array}{ccc}
x&y&0\\
y&-x&0\\
0&0&0\\
\end{array}\right).                            
\label{eq:11}
\ee
Then the interpretation is that we have a vortex with one-half
winding number centred on the $z$-axis and extending indefinitely
along it. To justify this interpretation, we have to exhibit an
Yang-Mills potential which will give rise to $T_{ij}$ as considered in
(\ref{eq:11}). It has been shown in Ref.\ \cite{ph} that in the generic
situation where the $3\times 3$ matrix $B_{i}^a$ is invertible and
smooth, there exists a smooth $A_{i}^a$ which will reproduce such a
$B_{i}^a$. So for the case here, $A_{i}^a$ can be constructed as a Taylor
series expansion about the origin. We will present such a series for a
different example below. We will also discuss the finiteness of energy
(per unit length) there.

We now show that monopoles of one-half winding number also occur.
The paradigm in this case is provided by the $3\times 3$ real symmetric matrix
\be
T=
\left(\begin{array}{ccc}
0&0&x\\
0&0&y\\
x&y&-2z\\
\end{array}\right)\,.
\label{thbyth}
\ee
Here the eigenvalues are $\l_{\pm}=-r(\cos\th \mp 1)$ and $\l_0=0$.
In the spherical co-ordinates, the corresponding eigenfunctions are
\be
\zeta^+=\left(\begin{array}{c}
\cos{\fr{\displaystyle\th}{\displaystyle 2}}\cos\phi\\
\cos{\fr{\displaystyle\th}{\displaystyle 2}}\sin\phi\\
\sin{\fr{\displaystyle\th}{\displaystyle 2}}
\end{array}\right),~~~
\zeta^-=\left(\begin{array}{c}
\sin{\fr{\displaystyle\th}{\displaystyle 2}}\cos\phi\\
\sin{\fr{\displaystyle\th}{\displaystyle 2}}\sin\phi\\
-\cos{\fr{\displaystyle\th}{\displaystyle 2}}\end{array}\right),~~~
\zeta^0=\left(\begin{array}{c}
-\sin\phi\\
\cos\phi\\
0
\end{array}\right).
\label{tcz}
\ee

\begin{figure}
\begin{center}
\includegraphics[scale=0.40]{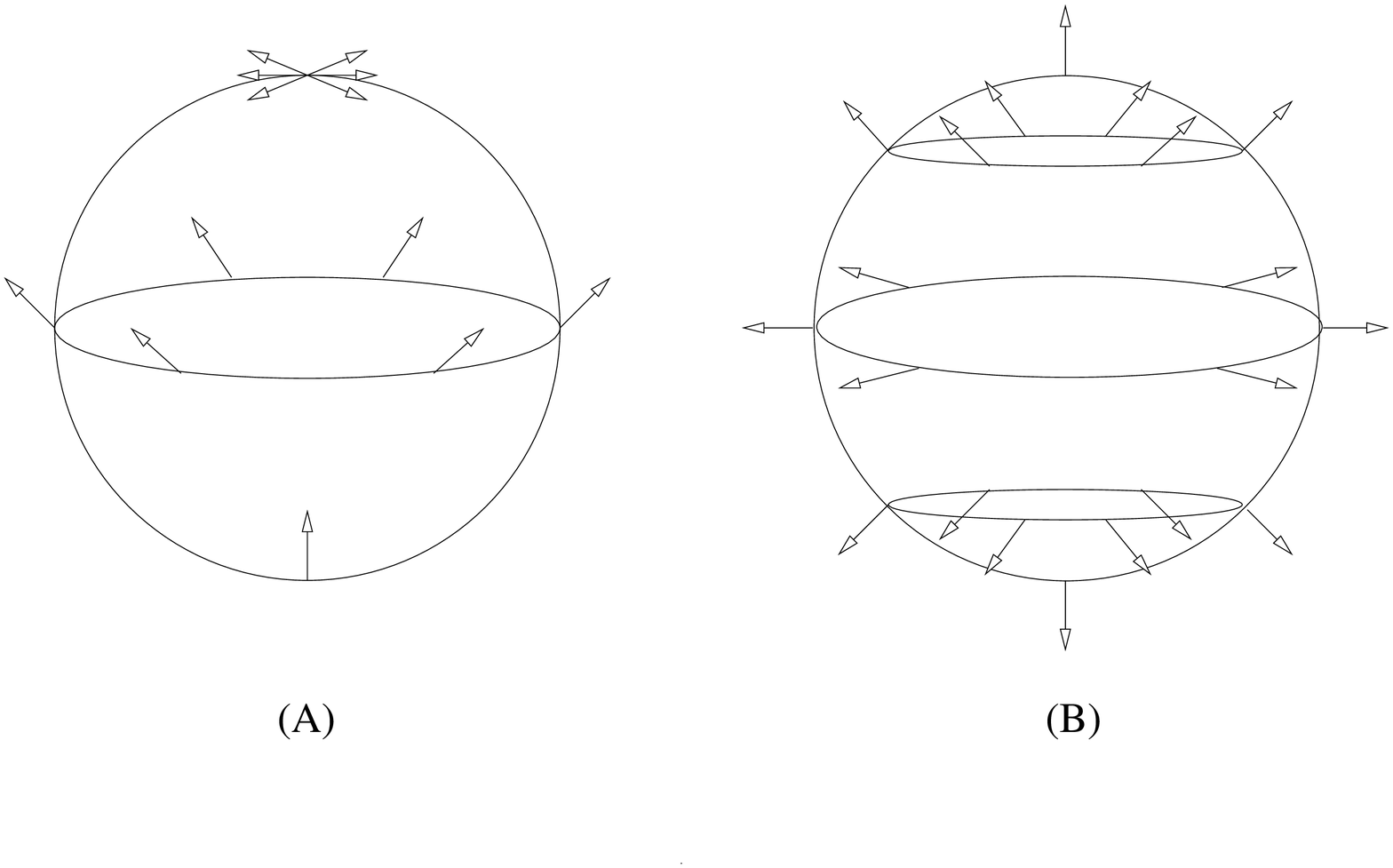}
\end{center}
\caption{(A) A winding number half configuration in three dimensions.
 There is a vortex of winding number one along the positive $z$-axis terminating
 at the centre. (B) A winding number one configuration in three dimensions. The
 upper half of this configuration is mapped onto the entire sphere in Fig.\ (A) to give
 one-half winding number.}
\label{fig3dim}
\end{figure}

Comparing  $\zeta_i^{\pm}$ with the radial vector and with the normalised
Higgs in the \TPM, viz. $(\sin\th \cos\phi, \sin\th\sin\phi,\cos\th)$,
we notice that essentially the angle $\th$ is replaced by $\th/2$.
This leads to one-half winding number in the present case.
This phenomenon is illustrated for $\zeta_i^+$ in Fig.\ \ref{fig3dim}. In effect, the
configuration in the upper half of the sphere for winding number one
is mapped onto the entire sphere to give one-half winding number. 
We refer to such a configuration as a half-monopole. Note
that the vector field $\zeta_i^+$ is singular (indeterminate in direction) all along the positive
$z$-axis. This is possible because $T_{ij}$ has a double degeneracy
there. This has the interpretation of a vortex (of winding number one)
along the positive $z$-axis terminating at the origin and giving rise
to a monopole. Because of this vortex, the vector field is not continuous on
the sphere, and therefore one-half winding number is possible. If the vector
field were smooth on the sphere, the winding number would have been only integral.

In an analogous way, $\zeta_i^-$ corresponds to a vortex
of unit winding number along the negative $z$-axis, terminating at
the origin. Finally $\zeta_i^0$ is again a vortex of winding number one
extending indefinitely along the $z$-direction.

That the monopole centre (point of triple degeneracy) is a terminating
point of vortex centre (line of double degeneracy) is a generic
situation. In fact, the generic situation is as follows. The
configuration $\zeta_i^A$, for each $A$, will have double degeneracy along 
two lines terminating at the centre. Each such line will be the centre
of a vortex of winding number half. This will be elaborated elsewhere.
It may also be noted that for the \TPM, which is not generic due to 
rotational invariance,
we have double degeneracy everywhere.

If we formally compute the Poincar\'e-Hopf index of the vector field
$\zeta_i^+$, we get it to be $-1/2$. The index for $\zeta_i^A$ is given by 
$M={\oint}_S dS^ik_i^A$,
where the integration is over a surface $S$ enclosing the centre and $k_i^A$
is the Poincar\'e-Hopf current \cite{his, goddard, arafune}
\be
k_{i}^A=\frac{1}{2}\e_{ijk}\e_{lmn}\zeta_{l}^A\p_j\zeta_{m}^A\p_k\zeta_{n}^A
~~~\mbox{(no sum over $A$)}\,.
\ee
We have in the present case (see Eq.\ (\ref{tcz}))
\be
k_i^+= -\hat x_i\frac{1}{4r^2}{\rm cosec}\,\frac{\theta}{2}\,.
\ee
The vector field $\zeta_i^+$ is not smooth at the north pole of
the sphere. Therefore, the definition
of the index $M$ is only formal. Nevertheless, this singularity is of zero
measure in the integration over $S$ and we get the winding number to
be $-1/2$. Note that the ``magnetic field" $k_i^+$ of this half-monopole is
not spherically symmetric, in contrast to the case of the Dirac monopole.
It has only an axial symmetry. 

In Ref.\ \cite{his}, it was shown that the Poincar\'e-Hopf current for the
eigenvector $\zeta_i^A$  can be expressed as the curl of an Abelian
vector potential $\w_i^A$:
\be
 k_{i}^A=\e_{ijk}\p_j\w_{k}^A ~- ~\mbox{Dirac string contributions}\,,
 \label{noDs}
 \ee
 where
\be
\w_{i}^A=\fr{1}{2}\e^{ABC}\zeta_j^B\p_i\zeta_j^C\,.
\ee
Here the indices $A$, $B$ and $C$, having the values 1,2 and 3, label
the three eigenvectors.
The Abelian vector potential corresponding to $\zeta_i^+$ is
\be
w_{i}^+=-{\hat \phi}_i\fr{1}{2r}\sec\fr{\th}{2}\,.
\label{pureg}
\ee 
This potential has the Dirac string along the negative
$z$-axis. This Dirac string is unphysical, in the sense that it does
not contribute to the ``magnetic field" $k_i^A$ (see Eq.\ (\ref{noDs})).
In contrast, the vortex line along the positive $z$-axis is physical,
and, because of it, the monopole does not have spherical symmetry.

Similarly, we get the Poincar\'e-Hopf index for $\zeta_i^-$ as
$-1/2$.
In the case of $\zeta_i^0$, notice that it spans a two-dimensional
vector space as we vary $\phi$. Therefore the index computed over $S$
will be zero (three-dimensional winding number is zero). On the other
hand, it makes sense to calculate the index over a two-dimensional
surface. For any such surface not containing the $z$-axis, we get
winding number one.

We now present the Taylor series expansion of $A_i^a$ about the origin which
leads to $T_{ij}$ considered in (\ref{thbyth}). Consider first the matrix
$(B)_{ia}=B_i^a$. In the symmetric gauge $(B)_{ia}=(B)_{ai}$ \cite{his},
we have $(B^2)_{ij}=S_{ij}$, so
that, for the case given in (\ref{thbyth}),
\bea
B=I+\frac{1}{2}
\left(\begin{array}{ccc}
0&0&x\\
0&0&y\\
x&y&-2z\\
\end{array}\right)+\cdots\, .                                    \label{eq:Bia}
\eea
Here the ellipsis indicates terms of higher order in the coordinates. The most general
Taylor expansion of $A_i^a$ about the origin is:
\bea
A_i^a=a_{ai}+b_{aij}x_j+c_{aijk}x_jx_k+\cdots\, .           \label{eq:Aia}
\eea
To obtain $B_i^a$ as given in (\ref{eq:Bia}), it suffices to take
\bea
a_{ai}=0,~~~b_{aij}=-\frac{1}{2}\epsilon_{aij}\,,\\
c_{aijk}=\frac{1}{2}(\epsilon_{ijp}M^a_{pk}+\epsilon_{ikp}M^a_{pj})\,, \label{eq:ccc}
\eea
where
\bea
M^1_{31}=-\frac{1}{6},~M^2_{32}=-\frac{1}{6},~M^3_{33}=\frac{1}{2}\,,    \label{eq:MMM}
\eea
and all other $M^a_{ij}$ are zero. Thus our solution for the gauge field is
\bea
A= \frac{1}{2}\left(\begin{array}{ccc}
-xy/3&z-y^2/3&-y+yz\\
-z+x^2/3&xy/3&x-xz\\
y&-x&0
\end{array}\right)+\cdots\,,
\eea
where $(A)_{ia}=A_i^a$.

We now address the question of finiteness of the energy of the
half-monopole, given by $E=\int d^3x\,S_{ii}/2$.
As $S_{ij}$ can be expanded in Taylor series about the origin, the energy
is finite in the ultraviolet. Also the infrared finiteness of the energy resides in the
scale factors of $S_{ij}$, such as $\a(r^2)$ and $\b(r^2)$ in Eq.\ (\ref{stpm}), and these
can be chosen appropriately to get a finite energy. 
Note that the one-half winding number is due to the tensorial structure of
$S_{ij}$, the eigenvectors $\zeta_i^A$ being unaffected by the scale factors.

In both two and three dimensions, we have seen that the phenomenon of one-half
winding number is due to the generic linear terms in the Taylor expansion of $S_{ij}$.
Nevertheless, there are crucial differences in the origin of this phenomenon in the two
cases. In two dimensions, the ambiguity in the sign of the eigenvector was the underlying
reason. The line of discontinuity (the ``branch cut") was arbitrary, except for the starting point.
In three dimensions, lines of double degeneracy terminating at the centre of the monopole were
necessary to give the necessary discontinuity in the form of a vortex. But these lines of
double degeneracy are rigid, in contrast to the branch cuts in two dimensions.

To conclude, we have pointed out in this letter that vortex and monopole configurations of one-half
winding number are present in the Yang-Mills theory. They arise from smooth Yang-Mills potentials,
and are indeed the generic configurations in contrast to the
\TPM.

{\it Note added in proof}. Some of the works which discuss vortices of half-integer
 winding number in other contexts are given in Ref.\ \cite{govaerts}.

\end{document}